# All-electrical detection of the spin-charge conversion in nanodevices based on SrTiO$_3$ two-dimensional electron gases

*Fernando Gallego[1], Felix Trier[1,2], Srijani Mallik[1], Julien Bréhin[1], Sara Varotto[1], Luis Moreno Vicente-Arche[1], Tanay Gosavy[3], Chia-Ching Lin[3], Jean-René Coudevylle[4], Lucía Iglesias[1], Félix Casanova[5,6], Ian Young[3], Laurent Vila[7], Jean-Philippe Attané[7] and Manuel Bibes\*[1]*

[1] Unité Mixte de Physique, CNRS, Thales, Université Paris-Saclay, 91767 Palaiseau, France.
[2] Department of Energy Conservation and Storage, Technical University of Denmark, 2800 Kgs. Lyngby, Denmark.
[3] Components Research, Intel Corp., Hillsboro, OR 97124, USA.
[4] Centre de Nanosciences et de Nanotechnologies, CNRS, Université Paris-Sud, Université Paris-Saclay, C2N, 91120 Palaiseau, France.
[5] CIC nanoGUNE BRTA, 20018 Donostia-San Sebastián, Basque Country, Spain
[6] IKERBASQUE, Basque Foundation for Science, 48009 Bilbao, Basque Country, Spain
[7] Univ. Grenoble Alpes, CNRS, CEA, SPINTEC, Grenoble, France.

The Magnetoelectric Spin-Orbit (MESO) technology aims to bring logic into memory by combining a ferromagnet with a magnetoelectric (ME) element for information writing, and a spin-orbit (SO) element for information read-out through spin-charge conversion. Among candidate SO materials to achieve a large MESO output signal, oxide Rashba two-dimensional electron gases (2DEGs) have shown very large spin-charge conversion efficiencies, albeit mostly in spin-pumping experiments. Here, we report all-electrical spin-injection and spin-charge conversion experiments in nanoscale devices harnessing the inverse Edelstein effect of SrTiO$_3$ 2DEGs. We have designed, patterned and fabricated nanodevices in which a spin current injected from a cobalt layer into the 2DEG is converted into a charge current. We optimized the spin-charge conversion signal by applying back-gate voltages, and studied its temperature evolution. We further disentangled the inverse Edelstein contribution from spurious effects such as the planar Hall effect, the anomalous Hall effect or the anisotropic magnetoresistance. The combination of non-volatility and high energy efficiency of these devices could potentially lead to new technology paradigms for beyond-CMOS computing architectures.

# 1. Introduction

A two-dimensional electron gas (2DEG) can appear at the interface between two materials with different electronic properties, typically a semiconductor and an insulator[1]. In this configuration, the electrons are confined to move in a two-dimensional plane parallel to the interface. The electronic properties of 2DEGs are highly dependent on the materials used to create the interface. Following their discovery and investigation in heterostructures based on Si or on III-V semiconductors [2,3], 2DEGs were realized in oxides, notably in SrTiO$_3$ (STO), a semiconductor with a band gap of 3.2 eV [1,4]. Remarkably, a 2DEG forms in STO when four or more unit cells (u.c.) of another wide band gap insulator such as LaAlO$_3$ (LAO) are grown on top of it. Later on, it was revealed that this 2DEG can in fact be formed by combining further materials with STO. Remarkably, the 2DEGs present a plethora of electronic properties, including high electron mobility[5], strong spin-orbit coupling (SOC)[6] and even superconductivity[7]. These properties offer opportunities for oxide 2DEGs to be used in many electronic devices, including field-effect transistors and spintronics devices[8]. Due to the asymmetric nature of their confining potential, oxide 2DEGs have broken inversion symmetry, and thus display a Rashba spin-orbit coupling (SOC) which can be harnessed to interconvert spin and charge currents[9,10].

The electronic structure of STO 2DEGs has been studied experimentally[11] and by DFT calculations[12]. In bulk STO, the 3d-orbitals degeneracy is lifted by the cubic crystal field, with the t$_{2g}$ levels having a lower energy than the e$_g$ levels. The t$_{2g}$ triplet forms the conduction band of STO. In STO 2DEGs, the perpendicular electric field created by the asymmetric confining potential lifts the degeneracy of the t$_{2g}$ levels. In this scenario, the light d$_{xy}$ band moves to lower energy compared to the heavier d$_{xz}$ and d$_{yz}$ bands[13]. At low filling, electrons typically lie in d$_{xy}$ sub-bands only, but as the carrier density increases, through back-gating for instance, a Lifshitz transition occurs when heavier sub-bands become populated[14].

The Rashba SOC present in these non-centrosymmetric 2DEGs is responsible for the two main physical effects that in low dimensions can transform a spin current into a charge current and vice-versa, the direct and inverse two-dimensional spin Hall effects[15] and the direct and inverse Edelstein effects (DEE and IEE)[16]. The DEE and IEE occur at the surface of topological insulators or at Rashba interfaces. In both these systems, the spin degeneracy is lifted, and at each point of the Fermi contours the spin is locked perpendicular to the electron momentum $\vec{k}$. Therefore[16], if one injects a charge current in the system, for example, along the $\vec{x}$ direction, the corresponding shift $-\Delta k_x$ of both Fermi contours will generate an excess spin density oriented along the $\vec{y}$ axis (DEE). The generated spin density can then diffuse to an adjacent region or an adjacent material, thus generating a pure spin current. Reciprocally, if a spin current is injected into a Rashba material, the two Fermi contours will be shifted in opposite directions, generating a transverse charge current along $-\vec{x}$ (IEE) [15]. The figure of merit used to estimate the efficiency of this conversion is the inverse Edelstein length $\lambda_{IEE}$, which is the ratio between the injected three-dimensional (3D) spin current and the generated 2D transverse charge current. This efficiency is proportional to the Rashba parameter $\alpha_R$ and the relaxation time $\tau$ [17].

Previous experiments have revealed a sizeable Rashba coupling in LAO/STO 2DEGs, up to $\alpha_R$ = 50 meVÅ [18,19]. Thanks to the relatively long relaxation time in STO 2DEGs ($\tau \approx 1$ ps), this Rashba coefficient is expected to translate into a large spin-charge conversion (SCC) efficiency through the IEE, and indeed $\lambda_{IEE}$ in the 1-10 nm range were measured at low temperature[9], exceeding the values obtained at Rashba interfaces and topological insulators [20,21]. $\lambda_{IEE}$ was also found to depend highly on the position of the Fermi level within the rich band structure of STO 2DEGs, tuned through back-gating. This is because the Rashba coefficient in $d_{xy}$ bands has a sign opposite to that of $d_{xz}$ and $d_{yz}$ bands. However, to date, SCC in STO 2DEGs has been mostly realized using spin-pumping, exploiting the ferromagnetic resonance of a ferromagnetic element to inject a pure spin current into the 2DEG. To assess further the potential of STO 2DEGs for spin-orbitronic devices, it is imperative to move towards the direct electrical injection of spins in nanoscale devices.

Among such spin-orbitronics devices, the Magneto-Electric Spin-Orbit (MESO) device[8] is a cutting-edge technology that has gained significant attention in recent years, due to its potential for beyond CMOS computing. The core of the MESO device is a ferromagnet (FM) used to store a binary information encoded in its magnetization direction. A key feature of MESO devices lies in the use of a voltage applied to a magnetoelectric input element to switch the magnetization of the FM and write binary information[22–24]. Very little current then flows through the device, making this operation intrinsically ultralow power. To read out the FM state, a spin-polarized current flows from the FM into the output section of the device involving a SOC material. The spin current is converted into a charge current, typically through IEE, generating a voltage, which may then be used to switch the next MESO element in cascaded structures. Essential to MESO operation is thus the magnitude of the output voltage, which depends directly on the SCC efficiency and the resistivity of the SOC system. Thanks to their SCC efficiency and large sheet resistance, STO 2DEGs show a high potential as the read-out block of the MESO device.

## 2. Results

In this work, we explore the potential of STO 2DEGs as the output unit of MESO devices. We fabricated 2DEG devices by growing different metals on top of STO//LAO (1.5 u.c.)[25], which eliminates the problem of increasing the temperature to grow the LAO at the end of the fabrication process, and strongly reduces the tunnel resistance for spin injection. We found a direct signature of inverse Edelstein effect by injecting a spin-current into the 2DEG and measuring the transverse signal. We show that this IEE signal is highly tunable by back-gating, consistent with earlier measurements using spin pumping[9,26]. Finally, we explored and were able to exclude different spurious contributions that may contribute to the measured signal.

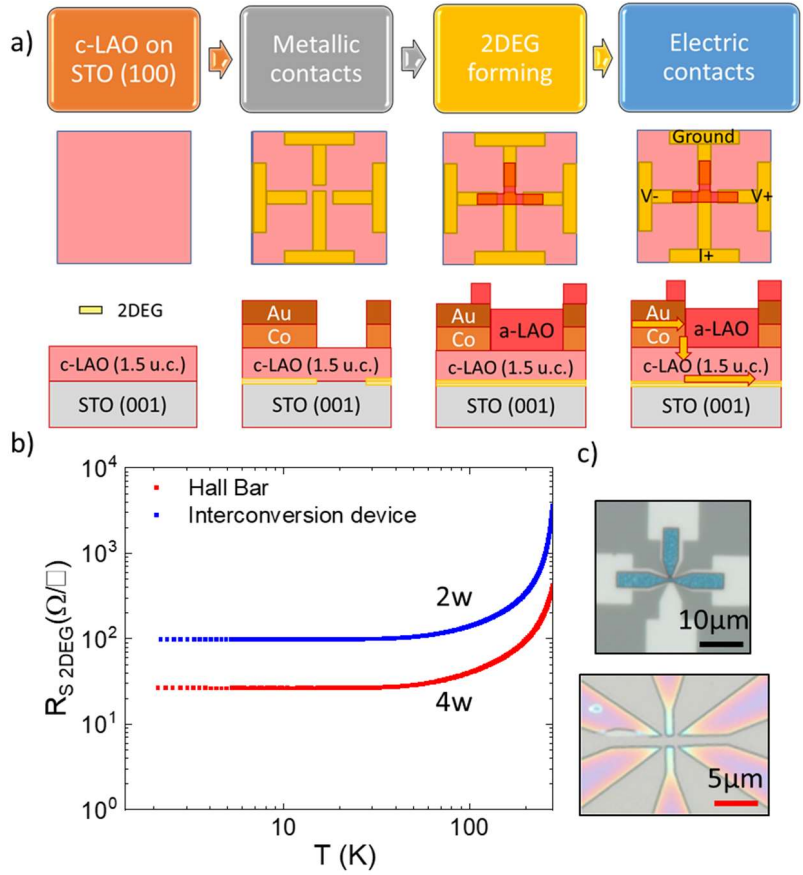

**Figure 1**: a) Flowchart of the device fabrication process. The first step consists of the growth of 1.5 u.c. of c-LAO on top of a TiO$_2$ terminated STO (001) substrate. The first lithography process is the design of four electrodes comprised of Au (3 nm)/ Co (18 nm) on the c-LAO. The second lithography step consist on growing the T-shape a-LAO in between the probes. b) Sheet resistance versus temperature for both a T-shaped device and a Hall bar formed by the same materials. The T-shaped device was measured in 2-probe configuration and the Hall bar was measured in 4-probe, both showing a qualitatively similar behavior. c) Optical microscope images of the two devices measured.

A sketch of the device fabrication process can be seen in **Figure. 1a**. We start by growing just 1.5 u.c. of crystalline LAO on the STO substrate, which is not thick enough to trigger the formation of the 2DEG[25]. This protects the interface where the 2DEG will be formed afterwards and maintains the high crystallinity of the surface [**Supplementary material 1**]. Once the electrodes are defined by electron-beam lithography and the Co layer is grown on top by sputtering, the presence of Co at the interface with the crystalline LAO (c-LAO) leads to the appearance of the 2DEG at the c-LAO//STO interface[25,27]. This approach limits the interface resistance for spin injection and allows tailoring the shape of the 2DEG without having to pattern the c-LAO layer. Similarly, a 2DEG is formed by depositing 30 nm of an amorphous LAO (a-LAO) layer on top of the c-LAO//STO at the intersection between the four Au/Co electrodes, creating a continuity that connects all the electrodes through the 2DEG only, as shown by the yellow arrows of the right panel of Figure 1a. The characterization of the 2DEG formed when amorphous LAO (a-LAO) is grown on c-LAO//STO can be seen in [**Supplementary material 2**].

During the process, Hall bars made of Au/Co/c-LAO (1.5 u.c.)//STO (001) were also fabricated in order to compare the electrical behavior of the 2DEG with the interconversion devices. Figure 1b displays the $R_{S\ 2DEG}$ vs. temperature measurement of both the device and the Au/Co Hall bar measured in 2-probe and 4-probe respectively. Except for a narrow region in the center of the device, the current always flows in parallel between the Au/Co metal part and the 2DEG, which is also the situation in the whole Hall bar. For both measurements, we thus extract the 2DEG contribution using[25,27] $\frac{1}{R_{device}} = \frac{1}{R_{2DEG}} + \frac{1}{R_{Metal}}$, and assuming that the resistance of the metal is constant with temperature, i.e. that $R_{metal} \approx R_{device}$ (300 K). The sheet resistance was then calculated for the Hall bar as $R_{S\ 2DEG} = R_{2DEG} \cdot W/l$, where W= 2 μm is the bar width and $l$ = 1.5 μm is its length from center to center of the electrodes. To calculate the sheet resistance of the device, we assume effective 4-probe measurement with $W_{FM}$= 300 nm, the width of the ferromagnetic channel of the device, and $l$ = 160 μm, which is the distance between the two transverse contacts. With this approximation, we see that, as expected, the 2DEG in both the device and the Hall bar shows the same general behavior, with a large decrease of the resistance upon decreasing temperature typical of 2DEGs[28]. The difference in the absolute sheet resistance values is likely due to contributions from the contact and leads series resistance in the device case.

To probe the presence of the IEE in the system, a charge current is injected from the Au/Co injector into the 2DEG (along the *y* axis) while we measure the transverse voltage (along the *x* axis) generated when the current flows from the Au/Co into the 2DEG, i.e. at the intersection of the four electrodes (Figure 1a). To respect the IEE symmetry, the magnetic field is swept along the *y* axis, such that the injected spins from the Co electrode are polarized along $\vec{y}$, and generate a charge current along $\vec{x}$. The measurement configuration can be seen in **Figure 2a**. The result for three different devices is shown in Figure 2 b, c and d. In these devices the dimensions and configuration of the electrodes are the same, while the width of the FM injector and 2DEG channel are varied, as shown in the insets of Figure 2 b, c and d. The applied back-gates for the loops in Figure 2 are +60 V (a), -60 V (b), and 0 V (c). The current injected from the Co into the 2DEG is spin-polarized, i.e. it is the sum of a pure charge current and a pure spin current (unlike in spin-pumping experiments where only a pure spin current is injected). The spin polarization at the Co/LAO interface was reported to be around 7% [29]. We expect the spin current to be converted into a transverse charge current in the 2DEG and to contribute to the output resistance $R_{Out}$, calculated as $R_{Out} = V_{Out}/I_{In}$. This transverse signal can also pick up charge-current based transverse voltages produced in the ferromagnet, due to the anomalous Hall effect (AHE), the planar Hall effect (PHE), the stray field out of plane at the tip of the ferromagnet, or the anisotropic magnetoresistance (AMR)[30], as will be discussed later. The signal shown in Figure 2 b, c and d was measured in AC in order to reduce the noise level [see **supplementary material 3**]. The clearest feature of the measurement is the presence of a hysteresis loop when the magnetization of the Co is reversed and, thus, the spins injected from the Co to the 2DEG flip. ΔR is calculated as the difference between the two resistance states in remanence. The amplitude of this jump ΔR varies from about 50 to 100 mΩ, proportionally to $W_{FM}$, the width of the FM channel, as expected from[31,32]. We can appreciate for the three different datasets that the loops are shifted towards negative magnetic fields, which should be related to exchange bias[33]. The value of this exchange bias is around 250 Oe. This shift happens naturally in magnetic systems that are in contact with an antiferromagnet and its magnetization gets pinned, shifting the coercive field. The generation of minute amounts of Co oxide at the interface between the Co and the c-LAO has been previously reported[25]. CoO is AFM with a Néel temperature $T_N$=280 K (in the bulk) and could thus be responsible for the exchange bias in the resistance loops.

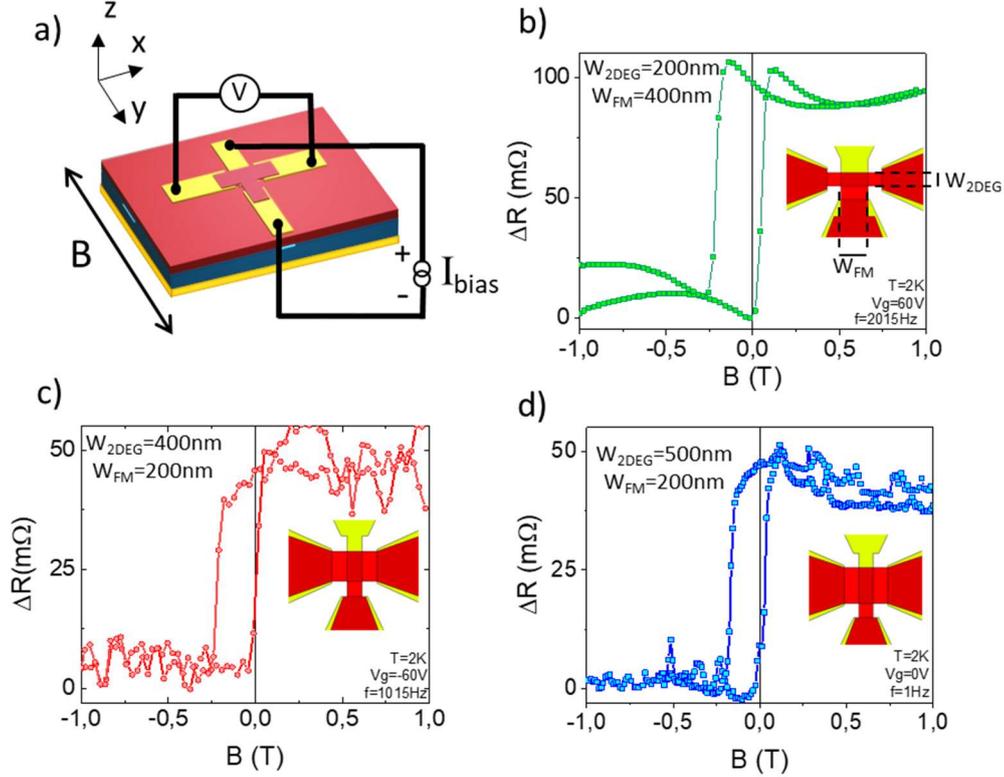

**Figure 2**: a) Sketch of the measurement configuration. b-d) Example of SCC feature in three different devices, all of them measured using the same configuration, by applying an AC current with a Keithley 6221 as current source and a Zurich HF2LI 50 MHz Lock-in Amplifier to measure the transverse voltage. Dimensions are shown for each device in the insets of (a-c). For clarity, a baseline resistance of 500-1500 mΩ was subtracted from the SCC loops.

Also, for the device in Figure 2b and slightly in Figure 2d, the data show dips and peaks at low magnetic fields. These features are usually related to the presence of PHE[32,34], that arises from a relative misalignment between the applied current and the magnetic field, and normally is caused by the experimental setup arrangement.

For the devices in Figure 2b (200×400 nm$^2$) and 2d (500×200 nm$^2$), a back-gating study was performed at T=2 K. The gate voltages were applied as shown in the inset of **Figure 3a**. For every applied gate, we performed a measurement from -1 T to +1 T, applying the current along the *y* axis and measuring the transverse voltage. In Figure 3a and 3c, the amplitude of the jump at remanence is represented as a function of the back voltage for the two devices, showing a non-monotonic evolution that has been previously reported for the same material in spin pumping experiments[26]. This peculiar dependence arises because the applied gate will shift the Fermi level through the complex band structure of the STO[35], with different bands having different Rashba coefficients. When only the $d_{xy}$ band is occupied, the Rashba coupling is positive and weak. When the Fermi level crosses the $d_{xz/yz}$ band, the Rashba coefficient changes its sign to negative. At this point, the SCC will thus behave nonmonotonically. In Figure 3b and 3d, two resistance loops are presented for two devices (200×400 nm² and 500×200 nm²), showing a difference of ΔR at remanence of 45% for the first one and 30% for the second, depending on the applied back-voltage.

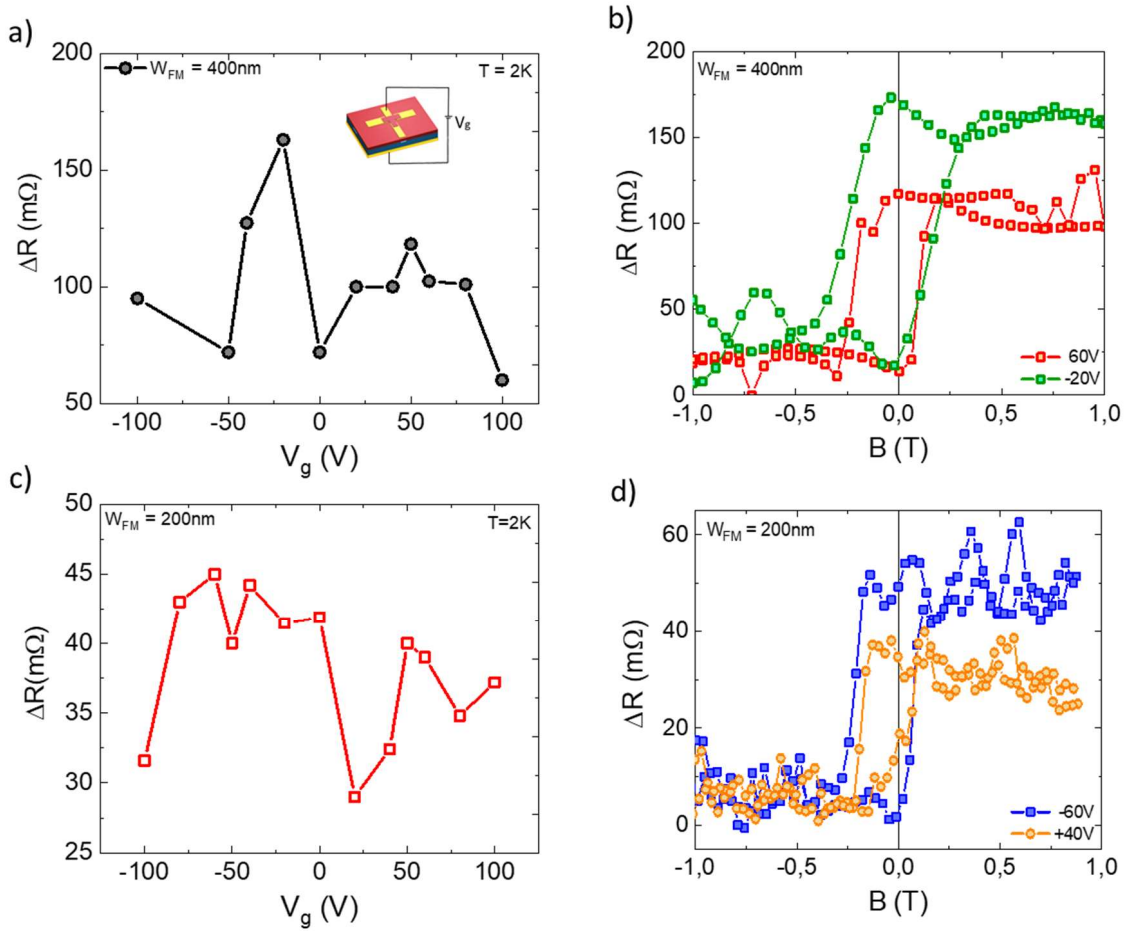

**Figure 3**: a,c) Evolution of spin-to-charge conversion amplitude as function of applied back-gating voltage at remanence of the cycles for the devices with dimensions 200×400 nm² (Figure

2b) and 500×200 nm² (Figure 2d). b,d) Examples of 2 SCC loops for -20 V and +60 V gate voltages applied in 200×400 nm² device and -60 V and +40 V for 500×200 nm² device.

By using the obtained values of the SCC amplitude, we can estimate the conversion efficiency ($\lambda_{IEE}$) of our devices. The efficiency is calculated as the 2D transversal charge current divided by the 3D injected spin current [see details in **supplementary material 4**]. Considering the output voltage of the $W_{2DEG}$=200 nm and $W_{FM}$=400nm device, the estimated conversion efficiency obtained is $\lambda_{IEE}$=0.72 nm. This is approximately one order of magnitude smaller than the one obtained from spin pumping experiments. This could be related to the 2DEG quality decrease after the fabrication process. The possible apparition of defects could compromise the conversion efficiency by decreasing the spin relaxation time. Furthermore, the $\lambda_{IEE}$ obtained comes from the transverse signal that carries the spurious effects of the charge current, plus the possible current backflow effect when the spin current reaches the second interface with the Co/Au electrode. Finally, the presence of disordered spins in the interfacial CoOx may induce spin flips and decrease the spin polarization of the injected carriers well below the nominal value of 7%.

In **Figure 4a**, the evolution of the SCC signal with temperature is studied for the device with the highest SCC amplitude (Figure 2b). Three observations can be made. First, the square hysteresis, due to the expected IEE, disappears above 200 K, as can be seen in Figure 4b. This can be explained by the relation between the $\lambda_{IEE}$ and the carrier mobility $\mu_e$ through the relaxation time, $\tau$ [**Equation 3**]: $\lambda_{IEE} = \frac{\alpha_R \tau}{\hbar} = \frac{\alpha_R m_e^*}{\hbar e} \mu_e$, where $\alpha_R$ is the Rashba coefficient of the 2DEG, $m_e^*$ is the effective mass of the band at the Fermi level, $\hbar$ the Planck constant, $e$ is the charge of the electron and $\tau$ is the spin relaxation time. In systems with broken inversion symmetry which, as STO 2DEGs, present Rashba interaction, the spin and momentum are locked. In the limit of sufficiently strong Rashba coupling, the momentum relaxation and the spin relaxation times are thus identical[36]. At high temperatures, the mobility decreases drastically, and the IEE vanishes (Figure 4c). We note that, together with its gate dependence, the decrease of the IEE with temperature allows ruling out that the signal comes from the ordinary Hall effect (OHE) of the 2DEG generated by the stray field from the Co[37], since the OHE of STO 2DEGs varies monotonously with the gate and hardly varies with temperature. The second remark corresponds to the exchange bias, mentioned above, that is present from 2 to 200 K. The last remarkable feature of the SCC loops is the presence of resistance dips at low negative magnetic fields, and peaks at low positive fields, as mentioned above. The first ones come from the PHE, and can be corrected just by rotating the relative angle between the current and the magnetic field, i.e. by rotating the device inside the measurement chamber. The peaks present in the hysteresis loop are coming from the AMR of the Co layer, which is the only signal remaining at room T [**see Supplementary material 5**].

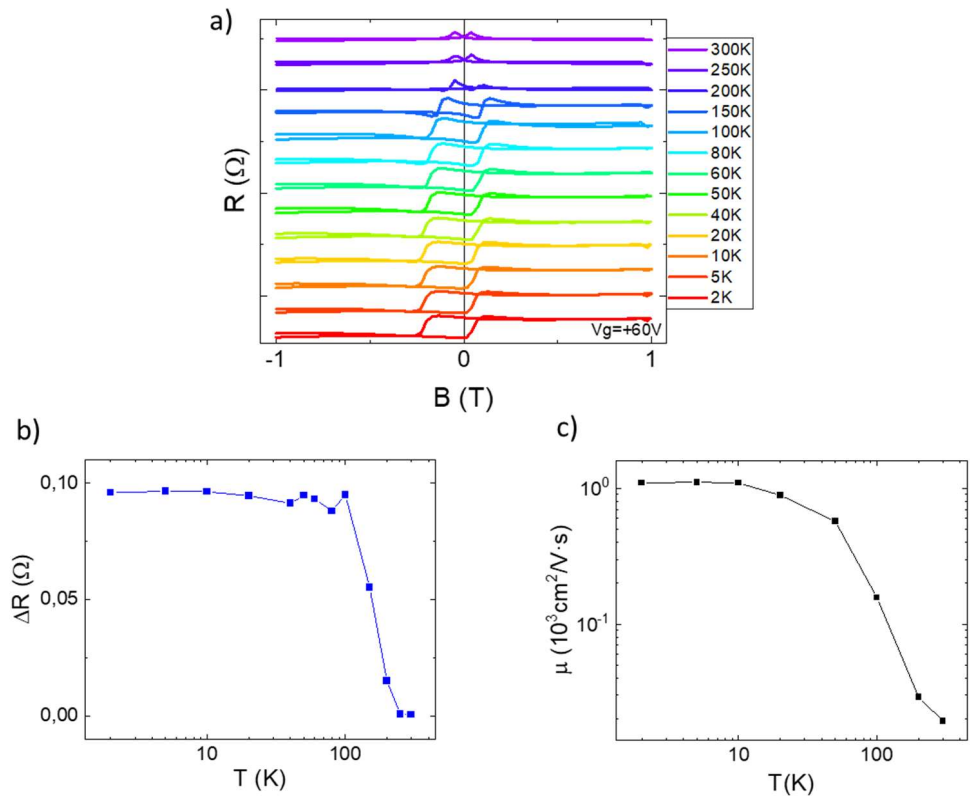

**Figure 4**: a) SCC evolution with the temperature for the device possessing the highest signal at Vg=+60 V. The loops were manually shifted by 100 mΩ for clarity. b) Loop amplitude vs. Temperature for the measurements at remanence. c) Evolution of the carrier mobility for the same temperature range in an a-LAO/c-LAO//STO thin film.

## 3. Conclusion

We have designed and fabricated nanodevices based on LAO/STO 2DEGs possessing a large spin-charge conversion efficiency. We have demonstrated the all-electrical spin injection and spin-charge conversion arising from the inverse Edelstein effect present in Rashba systems. Furthermore, thanks to the well-known tunability of the electronic structure of 2DEGs, we optimized the spin-charge conversion signal by applying back-gate voltages, and this has proven to be useful as a new degree of freedom. The optimized SCC signals are competitive with the values obtained for MESO's reading block by using metals or topological insulators at low temperature[32,34,37,38]. These results are encouraging in terms of spin-logic computing owing to the combination of magnetic non-volatility and a high energy efficiency. The gate tunability could open new possibilities for MESO and beyond.

## 4. Methods

*Sample growth and fabrication*:

A LaAlO$_3$ (1.5 u.c.) layer was grown by pulsed laser deposition (PLD) using a KrF laser ($\lambda$ = 248nm) on a TiO$_2$-terminated SrTiO$_3$ (STO) substrate (10×10×0.5 mm$^3$, miscut angle < 0.1°). The substrates were TiO$_2$-terminated by submerging them into an HF solution for 30 seconds. Subsequently, they underwent an annealing in O$_2$ atmosphere at 1000°C for 3 hours. The c-LAO was grown at 730°C and 2·10$^{-4}$ mbar of O$_2$, with a repetition rate of 1 Hz, an energy of 100 mJ, a fluence of 1.13 J/cm$^2$ and a target-substrate distance of approximately 5.5 cm. The samples were submitted to a one hour annealing at high O$_2$ pressure and 500°C. T-shaped devices were defined by using an Electron Beam Lithography (EBL) system and a lift-off process between steps. For all lithography processes the samples were cleaned in acetone (10 min) + IPA (10 min) before the spin coating of the PMMA A6 resist. The dose used for the EBL process was 300-600 μC/cm$^2$. The samples were developed for 2 minutes in a MIBK:IPA (1:3) solution, and followed by an $O_2$+Ar plasma cleaning process before every material deposition. Au and Co deposition were performed by conventional magnetron sputtering deposition at room T. The growth of 30nm a-LAO was performed in the PLD chamber after an in-situ $O_2$ plasma cleaning for 1 minute. After all the lithography processes, the outer parts of the devices were insulating with c-LAO (1.5 u.c.)//STO. Ti(10nm)/Au(50nm) back gate was sputtered on the bottom of the STO substrate by magnetron sputtering.

*Transport measurements:*

All the electrical measurements were acquired in a PPMS system from Quantum Design by using a Keithley 2400 sourcemeter for the DC measurements ($I_{DC}$ = 1-100 μA). For the AC measurements, a Keithley 6221 source ($I_{AC}$= 1-100 μA) was used in combination with a *Zurich HF2LI 50 MHz Lock-in Amplifier* ($f_{AC}$= 1 Hz - 2 kHz). A conventional Rotator PPMS puck was used in order to perform the in-plane measurements shown. Prior to any back-gate voltage studies, the LAO//STO devices went through a forming step at 2 K, where the back-gate voltage were cycled four times between −100 V and +100 V to ensure that no irreversible changes would occur in the interface system upon application of the back-gate voltage during the actual experiment. This back-gate contact was always grounded during cooldown experiments.

**Supporting Information**

Supporting Information is available from the Wiley Online Library or from the author.

**Acknowledgements**

This work received support from the ERC Advanced Grant No. 833973 "FRESCO, the ANR project CONTRABASS (ANR-20-CE24-0023-02), the European Commission under the H2020 FETPROACT Grant TOCHA (824140), and Intel's Science and Technology Center – FEINMAN. F.T. acknowledges support by research grant 37338 (SANSIT) from Villum

Fonden. F.C. acknowledges support by Spanish MICINN (Project No. PID2021-122511OB-I00 and Maria de Maeztu Units of Excellence Programme No. CEX2020-001038-M).

Received: ((will be filled in by the editorial staff))
Revised: ((will be filled in by the editorial staff))
Published online: ((will be filled in by the editorial staff))

Supporting Information

**All-electrical detection of the spin-charge conversion in nanodevices based on SrTiO$_3$ two-dimensional electron gases**

*Fernando Gallego, Felix Trier, Srijani Mallik, Julien Bréhin, Sara Varotto, Luis Moreno Vicente-Arche, Tanay Gosavy, Chia-Ching Lin, Jean-René Coudevylle, Lucía Iglesias, Fèlix Casanova, Ian Young, Laurent Vila, Jean-Philippe Attané and Manuel Bibes\**

**Supplementary material 1:**

Before the growth of every c-LAO sample, we confirmed by AFM that the atomic terraces were present at the substrate surface due to the chemical treatment and the annealing (Fig. S1a). The height of the atomic terraces is around 2Å (Fig. S1b). In Fig. S1c The RHEED pattern of the growth is presented, showing the 1.5 u.c. of c-LAO on top of the STO (100) substrate. The images in the insets of the figure show the evolution of the sample via a 2D growth.

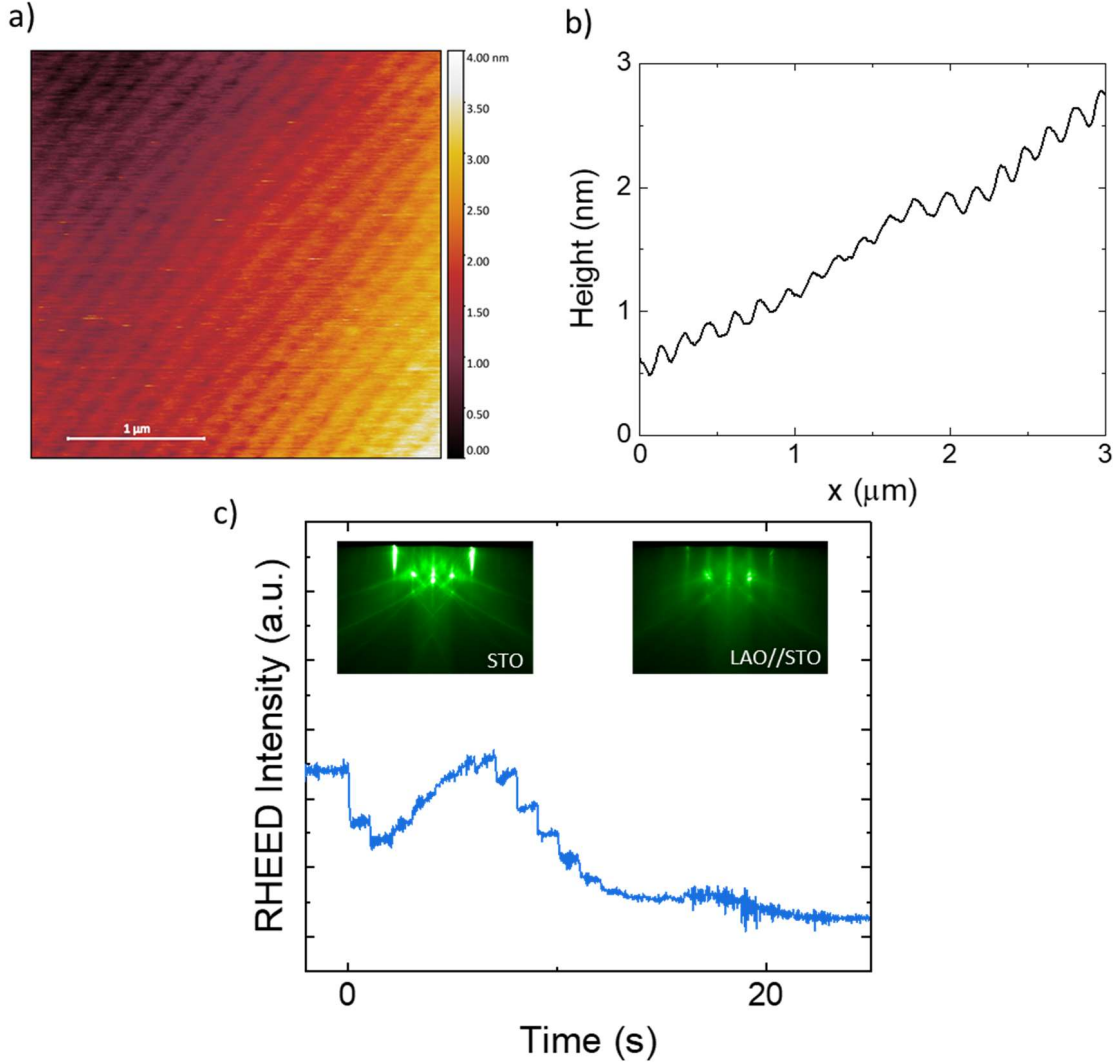

**Figure S1** : a) AFM image of a TiO$_2$-terminated STO (001) substrate. b) AFM profile of the atomic steps. c) RHEED oscillations for a c-LAO (1.5 u.c.) growth. The RHEED pattern of the substrate and the sample can be seen in the insets.

**Supplementary material 2:**

Thin-film samples of STO//c-LAO/a-LAO were grown in order to characterize the 2DEG created at the interface between the insulators. In Fig. S2a we see the metallic behavior of the gas, with a reduction of the resistance of 2 orders of magnitude. In Fig. S2b the magnetotransport data was measured in Hall configuration, showing a 2-band behavior at low temperatures. By fitting the Hall curves to a 2-band model, we can extract the carrier densities for the two types of carriers (Fig. S2c) and their respective mobilities (Fig. S2d). We can reach a carrier density of $2 \cdot 10^{13}$ cm$^{-2}$ with a high mobility reaching values of 10.000 cm$^2 \cdot$V$^{-1} \cdot$s$^{-1}$ for the minority carriers.

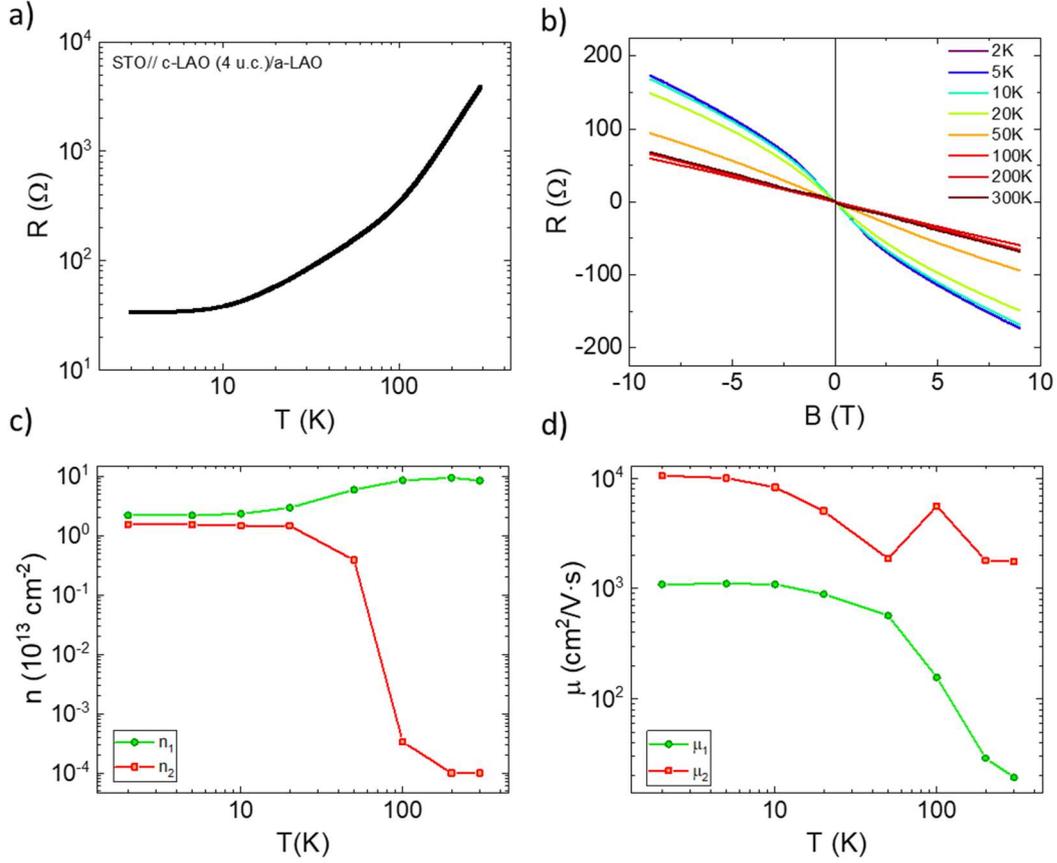

**Figure S2**: a) RvsT of a c-LAO (4 u.c.)/a-LAO sample grown on STO (001). b) Hall measurements at different temperatures. c) and d) carrier density and mobility at different temperatures extracted from the Hall measurements by fitting to a 2-band model.

**Supplementary material 3**

Here we present the difference in the measurement of devices in Hall configuration when measured in AC (a) or DC (b). In panel (a), the Resistance (X) and the Reactance (Y) are presented for the device in Fig. 2b from the main text, showing both a similar evolution with magnetic field, with Y being two orders of magnitude smaller. In (b), we can appreciate the same measurement taken in DC via a Keithley 2400 system.

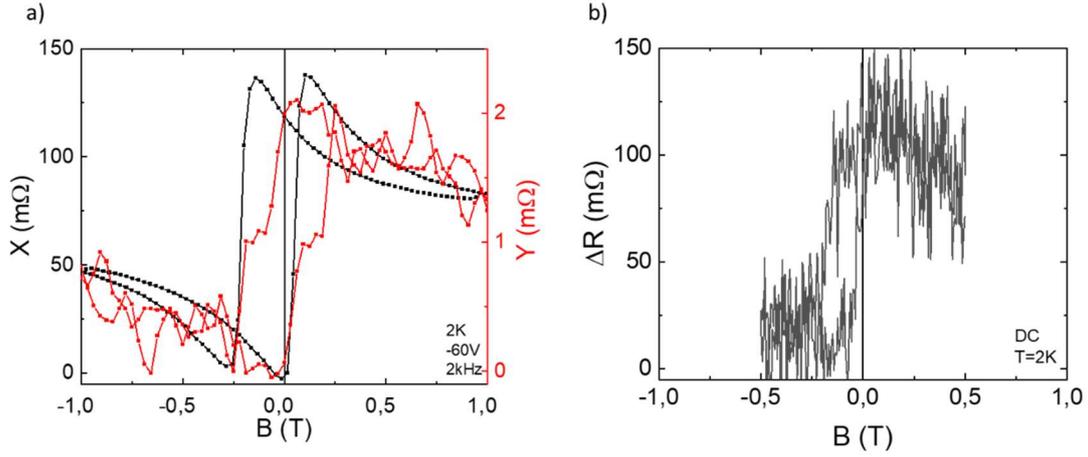

Figure S3 : a) Resistance (black) and Reactance (red) components of the AC measurement of the device from Figure 2.a. b) DC measurement of the same device.

**Supplementary material 4**

In order to calculate the $\lambda_{IEE}$ we took the SCC loop from the device with dimensions $W_{2DEG}$=200nm x $W_{FM}$=400nm. -the way of calculating the efficiency is as follows:

$$\lambda_{IEE} = J_{2D}/J_{3D}$$

$J_{3D}$ is the spin-polarized current that was injected into the 2DEG, calculated like:

$$j_{3D} = \frac{I_{In} \cdot P_{Co/LAO}}{A \cdot t}$$

Where $I_{IN}$=200µA is the applied current, $P_{Co/LAO}$ is the spin polarization at the Co/LAO interface, A=400x200nm² is the area of the Cobalt injection channel, and t is the thickness of the 2DEG. $J_{2D}$ is the generated transversal charge current. It is calculated as:

$$j_{2D} = \frac{V_{Out}}{R_{MR} \cdot S}$$

Where $V_{Out}$=15µV is the voltage amplitude of the SCC loop, $R_{MR}$=592Ω is the two-point resistance of the device at 2K, and S=200·t nm² is the section of the channel where the charge current will flow.

With these values we obtain a $J_{3D}$= 1.75/t ·10⁻¹⁰A/nm³ and a $J_{2D}$= 1.27/t ·10⁻¹⁰A/nm² and:

$$\lambda_{IEE} = 0.72 \; nm$$

**Supplementary material 5:**

In Fig. S5, we prove that we can, by rotating the sample from in-plane to out-of-plane, make the dips at low negative fields disappear. On the contrary, it is not possible to eliminate the contribution for the peaks by just rotating the sample while measuring. In Fig. S5b, the same measurement at 300K is presented. At this temperature there is no more contribution from the SCC, and only the peaks we appreciate at low temperatures remain here. This contribution could be coming from the anisotropic magnetoresistance of the Co layer.

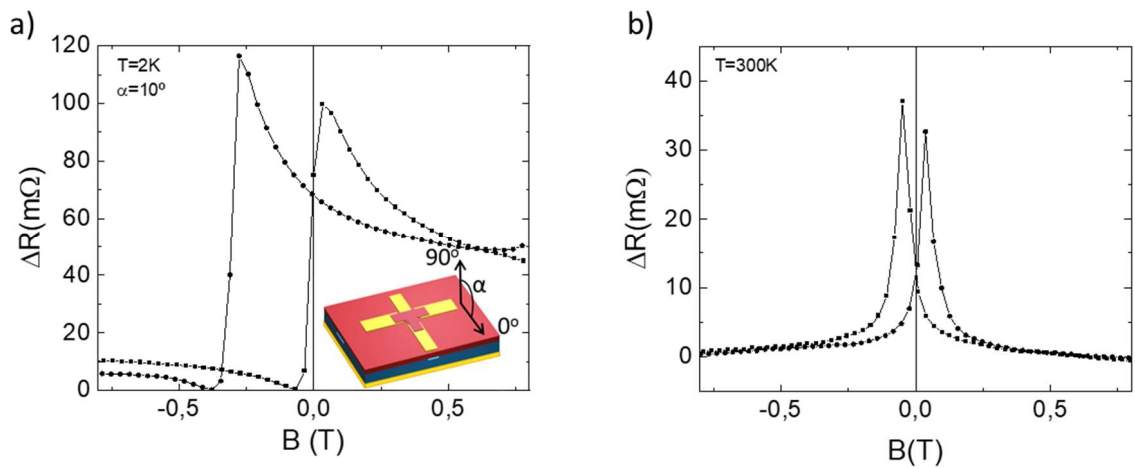

**Figure S5**: a) Spin-to-charge conversion signal of a T-shape device rotated from in-plane to out of plane configuration an angle α of 10°. The rotation angle α is represented in the inset. b) Remanence signal of SCC for the same device at 300K.